\documentclass[10pt]{article}
\usepackage{amsfonts, epsfig, amsmath, amssymb, color,mathabx,amsthm}
\usepackage{mathrsfs}

\usepackage[english]{babel}

\usepackage{mathrsfs}
\usepackage{pgfplots}
\pgfplotsset{width=7cm,compat=newest}
\usepackage{pgfplotstable}
\usepgfplotslibrary{statistics}
\usetikzlibrary{calc}

\newcommand{\E}{\mathbf{E}}
\renewcommand{\P}{\mathbf{P}}

\renewcommand {\epsilon}{\varepsilon}

\theoremstyle{definition}
\newtheorem{rem}{Remark}[section]
\newtheorem{exa}{Example}[section]

\DeclareMathSymbol{\ophi}{\mathalpha}{letters}{"1E}

\newcommand{\e}{\varepsilon}

\renewcommand{\phi}{\varphi}

\newcommand{\be}{\begin{equation}}
\newcommand{\ee}{\end{equation}}
\newcommand{\ben}{\begin{equation*}}
\newcommand{\een}{\end{equation*}}

\newcommand{\ba}{\begin{equation}\begin{aligned}}
\newcommand{\ea}{\end{aligned}\end{equation}}

\renewcommand{\i}{\mathrm{i}}
\newcommand{\ex}{\mathrm{e}}
\newcommand{\di}{\mathrm{d}}

\newcommand{\cA}{\mathcal{A}}

\newcommand{\cD}{\mathcal{D}}

\newcommand{\cN}{\mathcal{N}}

\newcommand{\cL}{\mathcal{L}}

\newcommand{\bI}{\mathbb{I}}

\newcommand{\bN}{\mathbb{N}}
\newcommand{\bR}{\mathbb{R}}

\usepackage{environ}

\newif\ifshowproof
\showprooffalse
\NewEnviron{Proof}{
    \ifshowproof
        \begin{proof}
            \BODY
        \end{proof}
    \fi
}

\newif\ifshowfigure
\showfigurefalse
\NewEnviron{Figure}{
    \ifshowfigure
        \begin{Figure}
            \BODY
        \end{Figure}
    \fi
}

\allowdisplaybreaks[4]

\let\oldmarginpar\marginpar
\renewcommand{\marginpar}[1]{\oldmarginpar{\scriptsize\texttt{\color{blue}{#1}}}}

\numberwithin{equation}{section}

\begin{document}

\title{How to simulate L\'evy flights in a steep potential:\\ An explicit splitting numerical scheme}

\date{\today} 


\author{Ilya Pavlyukevich\footnote{Institute of Mathematics, Friedrich Schiller University Jena, Inselplatz 5,
07743 Jena, Germany; ilya.pavlyukevich@uni-jena.de}, Olga Aryasova\footnote{Institute of Mathematics,
Friedrich Schiller University Jena, Inselplatz 5,
07743 Jena, Germany,
and Institute of Geophysics, National Academy of Sciences of Ukraine, Palladin Ave.\ 32, Kyiv
03680, Ukraine, and
Igor Sikorsky Kyiv Polytechnic Institute, Beresteiskyi Ave.\ 37, Kyiv 03056, Ukraine;
oaryasova@gmail.com}, Alexei Chechkin\footnote{Max Planck Institute of Microstructure Physics, Weinberg 2,
06120 Halle, Germany, and Wroclaw University of Science and Technology,
Faculty of Pure and Applied Mathematics, Wybrzeże Wyspiańskiego 27, 50-370 Wroclaw, Poland, and
Akhiezer Institute for Theoretical Physics, National Science Center ``Kharkiv Institute of Physics and Technology'',
Akademichna str.\ 1, Kharkiv 61108, Ukraine; achechkin@mpi-halle.mpg.de},
and Oleksii Kulyk\footnote{Wroclaw University of Science and Technology, Faculty of Pure and Applied Mathematics, Wybrzeźe Wyspiańskiego
27, 50-370 Wroclaw, Poland; kulik.alex.m@gmail.com}}

\maketitle

\begin{abstract}
We propose an effective explicit numerical scheme for simulating solutions of
stochastic differential equations with confining superlinear drift terms,
driven by multiplicative heavy-tailed L\'evy noise.
The scheme is designed to prevent explosion and accurately capture all finite moments of the solutions.
In the purely Gaussian case, it correctly reproduces moments of sub-Gaussian tails of the solutions.
This method is particularly well-suited for approximating statistical moments and other probabilistic characteristics of
L\'evy flights in steep potential landscapes.
\end{abstract}

\maketitle


\section{Introduction}

Non-Gaussian, heavy-tailed L\'evy processes have been a central focus of research across
various scientific fields over the past two decades. These are random jump processes with stationary
and independent increments, characterized by algebraically decaying probabilities of large deviations  ---
unlike Gaussian diffusions, which exhibit super-exponentially light tails.

One of the most prominent examples of such processes is the class of so-called
$\alpha$-stable L\'evy motions, also known as L\'evy flights, see \cite{UchaikinZ-99,nolan2020univariate}.
These are sometimes described as free anomalous diffusion.
They are characterized by
a continuous flow of tiny, incremental movements interspersed
with occasional large jumps that correspond to rare but significant (often catastrophic) events,
evoking a diffusive clustering behaviour punctuated by abrupt shifts.

Mathematically, $\alpha$-stable L\'evy motions emerge as scaling limits in the generalized central limit theorem
(see, e.g., Section 4.5 in \cite{whitt02}), or via random time changes (subordination)
applied to a free diffusion (see, e.g., Example 30.6 in \cite{Sato-99}). Analytically, they can be treated by means
of fractional calculus (see, e.g., \cite{MeerschaertS-12}).

In applications, heavy tailed statistics have been observed in
teletraffic \cite{resnick1997heavy,cappe2002long}, finance
\cite{rachev2003handbook,nolan2014financial},
biology \cite{koonin2006power},
physics \cite{shlesinger1995levy,chechkin2008introduction,schinckus2013physicists},
optimization \cite{kamaruzaman2013levy,chawla2018levy}, and
artificial intelligence \cite{simsekli2019tail,roy2021empirical}.

In the presence of external force fields, particularly external potential fields,
L\'evy flights exhibit more complex random dynamics.
Notable examples include overdamped L\'evy motion in periodic potentials (L\'evy ratchets) \cite{PavDybCheSok10,PavLiXuChe-15},
overdamped non-linear L\'evy oscillators \cite{ChechkinGKMT-02}, and L\'evy models of nonlinear friction
\cite{ChechkinGKM-05,KP-19}.
Of particular interest is the behavior of L\'evy flights in \emph{steep potentials} or systems with nonlinear dissipativity,
which has become an important topic of ongoing research, see, e.g.,
\cite{ChechkinGKM-04,DybGudHng07,dubkov2013time,garbaczewski2020levy}.
It is well known that steep
potentals can effectively \emph{confine} L\'evy flights. For example, whereas free L\'evy flights generally fail finite variance,
the variance of the confined motions becomes finite when subjected to sufficiently steep potentials. In particular,
for the quartic potential $U(x)\sim x^4$, the variance remains finite over an infinite time horizon for any stability
index $\alpha$,
see \cite{ChechkinGKM-04}.

The numerical approximation of heavy-tailed dynamics in steep potentials is of utmost importance.
However, many standard methods --- such as the classical Euler scheme (e.g., see Section 6.3 in \cite{JanickiW-94}) ---
often exhibit numerical instability or fail to accurately capture essential characteristics of the underlying random dynamics,
such as the variance or higher-order moments.

In this note, we introduce a novel explicit numerical scheme for the efficient simulation of confined, L\'evy-driven
heavy-tailed dynamics, which overcomes these deficiencies.

The paper is organized as follows. We begin with an illustrative example of a numerial approximation
for Cauchy noise in a steep potential $U(x)=x^{10}/10$.
In Section \ref{s:euler}, we discuss the limitations of the widely used Euler scheme and
highlight its deficiencies in capturing key properties of the random dynamics.
In Section \ref{s:splitting}, we introduce the direct and reverse splitting schemes for this class of equations.
Through numerical experiments, we show that the direct splitting scheme is the only method that is numerically stable and
accurately reproduces the moments of the original process.
In Section \ref{s:linear}, we perform a comparative error analysis of the Euler, direct and and reverse
splitting schemes for a linear system driven by symmetric $\alpha$-stable L\'evy noise.
Section \ref{s:general} contains the formulation of the direct splitting scheme in a general setting.
In Section \ref{s:examples}, we provide a collection of ready-to-use examples that can be employed
to simulate a variety of widely used nonlinear systems.
Finally, Appendix \ref{a:general} lists
the set of rigorous mathematical assumptions under which convergence of the direct splitting method has been established in
\cite{aryasova2025tail}.
Appendix \ref{a:simulation} describes several methods for simulating increments of heavy-tailed L\'evy processes.

\section{Cauchy noise in a steep potential: Euler schemes\label{s:euler}}

Let $X$ be the solution to the one-dimensional SDE
\ba
\label{e:exa}
\di X_t=-X^9_t\, \di t + \di Z_t, \quad X_0=x,
\ea
where the drift term $A(x)=-x^9$ is a (minus) gradient of a steep confining potential $U(x)=x^{10}/10$.
The driving noise $Z$ is a Cauchy process, a heavy-tailed L\'evy process with characteristic function
$\E \ex^{\i \lambda Z_t} =\ex^{-t|\lambda|}$, $\lambda\in \bR$.
Existence and uniqueness of solutions to this SDE follow from well-known results in the theory of L\'evy-driven SDEs; see, e.g.,
\cite{Applebaum-09}.

\begin{figure}
\centerline{\includegraphics{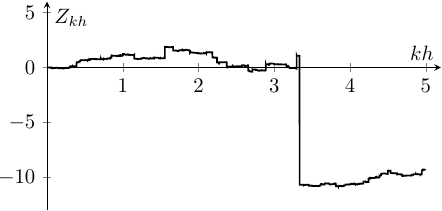}}
\caption{A sample path of a Cauchy process $Z$ on the time grid $\{kh\}_{k\in\bN_0}$, $h=10^{-3}$.\label{f:Z}}
\end{figure}

The Cauchy process $Z$ is a purely discontinuous L\'evy process
with heavy tails. It is integrable only up to order $q=1$: specifically, for $q\in(0,1)$
\[
\E |Z_t|^q= t^q (\cos(\pi  q/2))^{-1},
\]
while for $q\in[1,\infty)$, we have $\E|Z_t|^q=+\infty$.
In particular, the Cauchy process lacks both a finite mean and variance.
A typical sample path of $Z$, shown in Fig.~\ref{f:Z}, is characterized by rare but pronounced large jumps.

On the other hand, it is known, see, e.g., \cite{ChechkinKGMT-03,SamorodnitskyG-03,ChechkinGKM-04,ChechkinGKM-05,KP21}
that the fast growing confining drift $A$
has a \emph{confining} effect on the heavy-tailed noise.
Specifically, the solution $X$ of \eqref{e:exa} possesses all moments up to order $q=9$, uniformly bounded in time.
That is, for any $q\in [0,9)$,
\[
\sup_{t\in[0,\infty)}\E |X_t|^q<\infty.
\]
Moreover, the threshold $q=9$ is sharp: for every $t\in(0,\infty)$ and $q\in[9,\infty)$
\[
\E |X_t|^q=+\infty.
\]
This phenomenon --- where the solution exhibits finite moments beyond the integrability of the driving process $Z$ ---
is known as the \emph{confinement} or \emph{tail-improving effect}.

Although the moments $\E |X_t|^q$ cannot, in general, be determined analytically for a fixed $t\in (0,\infty)$,
their asymptotic values as $t\to\infty$ can be obtained.
Specifically, as a consequence of the superlinear confinement due to the
drift,
the distribution of $X_t$ converges exponentially fast to a stationary distribution $X_\infty$ as $t\to\infty$.

\begin{figure}
\centerline{\includegraphics{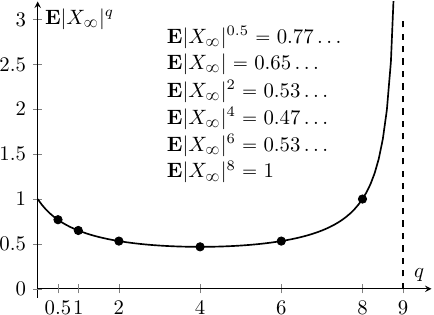}}
\caption{The absolute moments of the limit distribution $\E |X_\infty|^q$, $q\in[0,9)$.\label{f:moments}}
\end{figure}

The law of $X_\infty$ is known explicitly and has the density
\ba
\label{e:m}
m_\infty(x)
=\frac{1}{\pi(1+x^2)}\frac{1}{(1-2 x^2 \cos(\frac{ \pi}{9})+x^4)(1-2 x^2 \cos(\frac{5 \pi}{9})+x^4)},\quad x\in\bR,
\ea
see Eq.\ (38) in \cite{DubSpa07}.
This formula allows for the exact computation of the absolute moments of $X_\infty$.
Fig.~\ref{f:moments} illustrates these stationary moments, with specific numerical values given for moments of orders
$q=0.5$, 1, 2, 4, 6, and 8.

A commonly used approach for numerically solving L\'evy-driven SDEs is the
straightforward explicit Euler scheme on an equidistant time grid $\{kh\}_{k\in\bN_0}$; see, e.g.,
\cite{ChechkinGKM-cross-05,ChechkinGKM-04,delGonChe08,DybGudHng07,DybGud09,PavDybCheSok10}.
For our example \eqref{e:exa}, the explicit scheme $X^{\mathrm{E},h}$ is defined
by the recursive relation
\ba
\label{e:E-exa}
X^{\mathrm{E},h}_{0}&=x,\\
X^{\mathrm{E},h}_{(k+1) h}&=X^{\mathrm{E},h}_{kh} - ( X^{\mathrm{E},h}_{kh} )^9 h
+ Z_{(k+1)h} - Z_{kh},\quad k\in \bN_0.
\ea
The heavy tail increments in \eqref{e:E-exa} can be efficiently simulated as i.i.d.\
Cauchy random variables scaled by the time step $h$. Specifically,
\[
Z_{(k+1)h} - Z_{kh}\stackrel{\di}{=}h\xi_{k+1},\quad k\in \bN_0,
\]
where $\{\xi_k\}_{k\in\mathbb N}$ are i.i.d.\ standard Cauchy distributed random variables with the probability density
$p(x)=\frac{1}{\pi}\frac{1}{1+x^2}$.

Practical simulations of sample paths of $X^{\mathrm{E},h}$ reveal a significant drawback of the Euler scheme --- namely,
its susceptibility to numerical instability. This issue is not unique to L\'evy-driven systems and is known to occur
in the Gaussian case, as well; see, e.g., \cite{petersen1998general}.
Due to the unbounded nature of the noise increments and the superlinear growth of the drift,
there is a positive probability that the numerical approximation explodes,
leading to divergence in finite time.
In such scenarios, the scheme typically terminates with a NaN (``Not a Number") value,
indicating numerical blow-up.

\begin{figure}
\centerline{\includegraphics{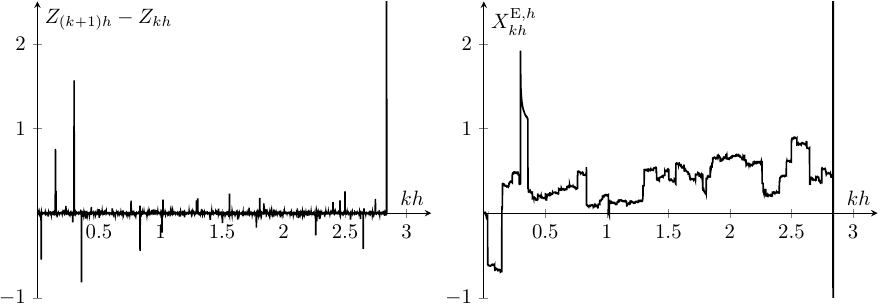}}
\caption{The blow up of the explicit Euler scheme after a large jump at time $k_*h=2.837$; $h=10^{-3}$, $x=0$.\label{f:blowup}}
\end{figure}

Let us provide a quantitative explanation of the blow-up phenomenon using a concrete example; see Fig.~\ref{f:blowup}.

We consider the  Euler scheme $X^{\mathrm{E},h}$ with the initial value $x=0$ and the step size $h=10^{-3}$.
We observe that at time instant $t_*=k_*h=2.837$, $X^{\mathrm{E},h}_{k_*h}=0.416$,
and the noise increment takes a relatively large value
$h\xi_{k_*+1}=4.9130$. This triggers a blow-up within three subsequent steps:
\begin{equation*}
\begin{aligned}
X^{\mathrm{E},h}_{k_*h}      &=0.416, \quad   &   h\xi_{k_*+1}&=4.9130,\\
X^{\mathrm{E},h}_{(k_*+1) h} &= 5.329, \quad & h\xi_{k_*+2} &=-0.001,\\
X^{\mathrm{E},h}_{(k_*+2) h} &= -3460.5, \quad & h\xi_{k_*+3}& =0.0009,\\
X^{\mathrm{E},h}_{(k_*+3) h} &= 7.1\cdot 10^{28}, \quad & h\xi_{k_*+4}& =-0.0004,\\
X^{\mathrm{E},h}_{(k_*+4) h} &= -\infty. &&
\end{aligned}
\end{equation*}
In other words, the numerical sign alternating instability develops rapidly once the current value $x=X^{\mathrm{E},h}_{(k^*+1)h}$
exceeds a certain threshold $M_h$. This threshold can be estimated via a
simple one-step analysis.

Assume, for definiteness, that $X^{\mathrm{E},h}_{(k^*+1)h}=x\in(0,\infty)$.
For instability to occur, the next iterate
$X^{\mathrm{E},h}_{(k^*+2)h}$ must change sign and have a larger magnitude than $x$.
 That is, the following inequality must be satisfied:
\ba
\label{e:x}
x - x^9h + h \xi_{k^*+2}< -x.
\ea
Since the noise typically does not take large values over two consecutive steps, the noise increment
$h \xi_{k^*+2}$ is of order $h$ with high probability and can thus be neglected.
Solving the inequality \eqref{e:x} gives a rough estimate for the instability threshold:
\[
|x|>(2h^{-1})^{1/8}=M_h.
\]
Because of confinement and symmetry, we can also assume that $X^{\mathrm{E},h}_{k^*h}\approx 0$.
In summary, the process $X^h_{(k^*+1)h}$ crosses
the instability level $M_h$ whenever
the noise increment $h\xi_{k^*+1}$ satisfies $h|\xi_{k^*+1}|>M_h$.

\begin{figure}
\centerline{\includegraphics{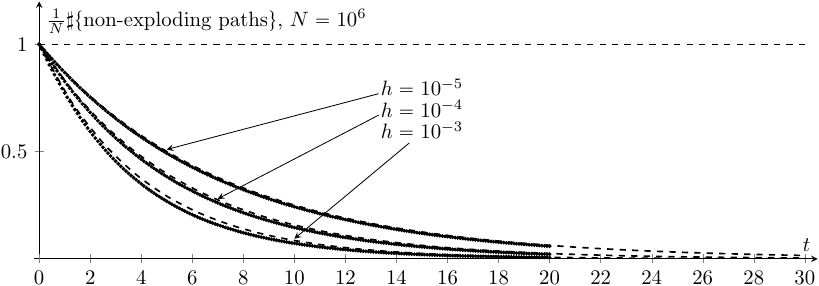}}
\caption{Relative number of stable paths of $X^{\mathrm{E},h}_{kh}$, $kh\leq t$,
for the explicit Euler scheme for $h=10^{-3}$,
$h=10^{-4}$, $h=10^{-5}$, $N=10^6$, $x=0$,
$t\in[0,20]$, and their approximations \eqref{e:stable} on the time interval $t\in[0,30]$ (dashed).
\label{f:NAN}}
\end{figure}

Consequently, the probability that the Euler scheme
$X^{\mathrm{E},h}$ remains
\emph{stable} over the interval $[0,t]$ can be estimated by
the probability that the noise increments does not exceed $M_h$ for all $kh\leq t$:
\ba
\label{e:stable}
\P(|X^{\mathrm{E},h}_{kh}|<M_h,\ kh\leq t)
& \approx \P(|h\xi_k|< M_h,\ kh\leq t)\\
& \approx \Big(1-\P(|\xi|> M_h/h)\Big)^{t/h}\\
& \approx \Big(1-\frac{2h}{\pi M_h }\Big)^{t/h}\\
&\approx \ex^{-\frac{2}{\pi M_h}t}.
\ea
This simple argument shows close agreement with the empirical data.
We simulate $N=10^6$ independent realizations of the Euler scheme $X^{\mathrm{E},h}$
on the interval $[0,t]$, $t= 20$, for several
time steps $h$. The relative proportion of non-exploding paths is shown in Fig.~\ref{f:NAN}.
For $h=10^{-3}$, $h=10^{-4}$, $h=10^{-5}$ the corresponding instability thresholds $M_h$ are
$2.6$, $3.45$ and $4.6$, respectively.
The theoretical prediction \eqref{e:stable} matches
the empirical data very well, particularly for smaller values of
$h$.

\begin{figure}
\includegraphics{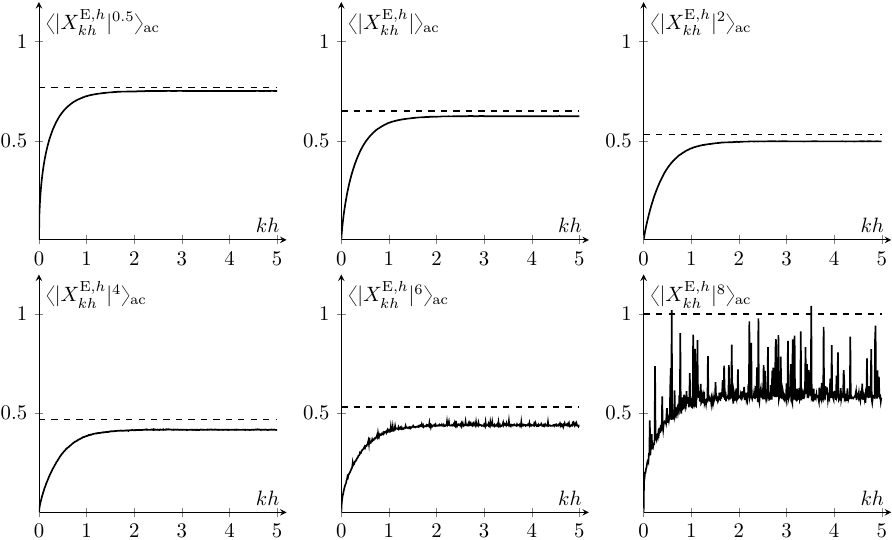}
\caption{``Available-case'' empirical absolute moments of the Euler scheme,
$h=10^{-5}$, $N=10^6$, $x=0$, calculated from $N_\mathrm{ac}\approx 5\cdot 10^5$ non-exploding trajectories on the interval $kh\in [0,5]$.
The dashed lines represent the corresponding moments of the limit distribution $X_\infty$.
\label{f:ac}}
\end{figure}

Despite the NaN problem, the Euler scheme remains widely used in simulations,
often with NaN outcomes simply disregarded.
In this context, one considers the statistics of ``observed values" or ``available-case" scenarios.
In Fig.~\ref{f:ac}, we present the
``available-case'' empirical absolute moments computed from $N=10^6$ independent runs of the Euler scheme.
They exhibit the following behaviour: the moments $kh\mapsto \langle|X_{kh}^{\mathrm{E},h} |^q\rangle_\mathrm{ac}$
stabilize at some level for large $t=kh$,
but this level is significantly smaller than the true moments, especially for higher orders $q$.

This effect can be explained by the fact that the non-exploding ``available-case'' paths
are effectively driven by bounded noise increments that do not exceed the effective threshold $M_h$.
As a result, the moments are substantially underestimated.

The blow-up challenge can be overcome by applying of the so-called tamed Euler scheme, that truncates the drift:
\ba
\label{e:TE-exa}
X^{\mathrm{TE},h}_{0}&=x,\\
X^{\mathrm{TE},h}_{(k+1) h}&=X^{\mathrm{TE},n}_{kh} - \frac{( X^{\mathrm{TE},n}_{kh} )^9}{1+ |X^{\mathrm{TE},n}_{kh}|^9 h} h
+ Z_{(k+1)h} - Z_{kh},\quad k\in \bN_0,
\ea
as suggested in \cite{hutzenthaler2012strong} for Gaussian SDEs, see modifications
for the L\'evy case in \cite{dareiotis2016tamed,kumar2016tamed}.

\begin{figure}
\includegraphics{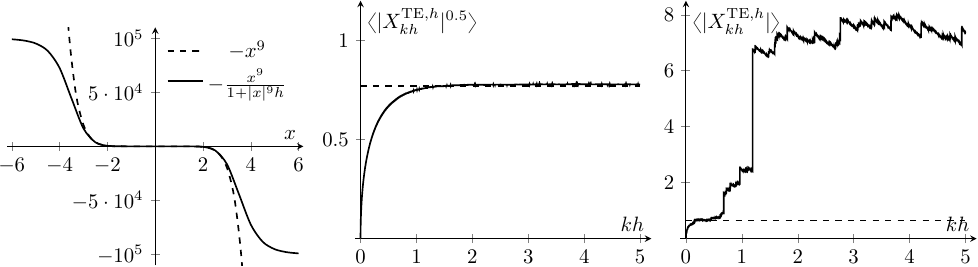}
\caption{Left: The drifts of the Euler and Tamed Euler schemes.
Middle and right:
Empirical absolute moments of the Tamed Euler scheme, $h=10^{-5}$, $N=10^{-6}$, $x=0$.
The dashed lines represent the corresponding moments of the limit distribution $X_\infty$.\label{f:TE}}
\end{figure}
In this scheme, the non-linear drift $A(x)=-x^9$ is replaced by the truncated version
$\frac{A(x)}{1+|A(x)|h}=-\frac{x^9}{1+|x|^9 h}$, which is bounded by $h^{-1}$, see Fig.~\ref{f:TE} (left),
ensuring that the scheme does not explode.
However, this truncation obviously removes the system’s confinement, which in turn affects the accuracy of the final results.
In Fig.~\ref{f:TE} (middle and right),
we present the empirical absolute moments of the tamed Euler scheme. For $q=0.5$ and sufficiently
large times $kh\geq 3$, the empirical moments converge to the true stationary value $\E |X_\infty|^{0.5}$,
as predicted in \cite{dareiotis2016tamed,kumar2016tamed}. For $q=1$, however, the moments are strongly overestimated.
For $q=2$, the moments stabilize around approximately $7500$ and are therefore not shown.

We do not discuss step-adaptive or implicit numerical schemes here. To the best of our knowledge,
convergence of moments for these schemes has not yet been studied in the context of heavy-tailed noise.

\section{Cauchy noise in a steep potential: Splitting schemes\label{s:splitting}}

To overcome the difficulties outlined above,
we propose an explicit \emph{splitting} numerical scheme for simulating solutions of \eqref{e:exa} that respects all the
moments of the true solution.

Our scheme is based on the well-known Lie--Trotter type approximations for semigroups, see, e.g.,
Section VIII.8 in \cite{reed1980methods} and
\cite{mclachlan2002splitting} for the general theory.
The idea behind this method can be explained as follows.
The process $X$ is a Markov process with generator
\[
\mathcal L f(x)= \cA f(x) +\cD f(x),\\
\]
where $\cA$ stands for the drift part
\[
\cA f(x)= A(x) f'(x)= -x^9 f'(x)
\]
and
$\cD$ stands for the noise part, in our example,
\begin{equation*}
\begin{aligned}
\cD f&= -(-\Delta)^{1/2} f(x)=\frac{1}{2\pi}\int_0^\infty \frac{f(x+y)-2f(y)+f(x-y)}{y^2}\, \di y.
\end{aligned}
\end{equation*}
The semigroup (propagator) $P$ of $\cL$ is expressed in terms of the process $X$ started at $X_0=x$ as
\[
P_t f(x)=\E f(X_t(x)) = \ex^{t\cL}f(x),
\]
whereas the propagators of $\cA$ and $\cD$ can be calculated directly, namely,
\[
\ex^{t \cA} f(x) =f(\Phi(t,x)),\\
\]
where $\Phi=\Phi(t,x)$ is the solution of the ODE
\ba
\label{e:ODE9}
\dot \Phi(t,x)&=A(\Phi(t,x)),\\
\Phi(0,x)&=x,
\ea
and
\[
\ex^{t \cD} f(x) = \E f(x+Z_t)=\int_{-\infty}^\infty \frac{1}{\pi} \frac{t}{t^2 + z^2}f(x+z)\,\di z.
\]
Then, the Lie--Trotter approximation theorem (the Lie--Trotter product formula)
states that for each $t\in[0,\infty)$
\[
\ex^{t \cL }= \ex^{t(\cA+\cD)}
= \lim_{h\to 0} (\ex^{h\cA}\ex^{h\cD })^\frac{t}{h},
\]
see, e.g., Section 2.12 in \cite{varadarajan2013lie}, Section 5.3 in \cite{Kolokoltsov-11}
or Section VIII.8 in \cite{reed1980methods}.

In other words, by taking $h$ small enough such that $t=kh$, we can approximate the propagator of $X$ by the
$k$-fold
composition of propagators of $\cA$ and $\cD$:
\ba
\label{e:P}
P_t \approx P^h_t=(\ex^{h\cA }\ex^{h\cD})^k.
\ea
Equation \eqref{e:P} provides an approximation of $X$ in terms of semigroups, i.e., in the weak sense.

We propose an analogue of the Lie--Trotter approximation \eqref{e:P} formulated directly
in terms of the processes, i.e., in the strong sense.

More precisely, we split the dynamics $X$ into two parts: one following the integral curves of the ODE
$\dot\Phi=A(\Phi)$
and the other given by the noise $Z$.
In our example, since $A(x)=-x^9$, the ODE
\eqref{e:ODE9}
takes the form $\dot \Phi= -\Phi^9$, which can be solved explicitly using the method of separation of variables. Rewriting
\eqref{e:ODE9} as
\[
-\frac{\di \Phi}{\Phi^9}= \di t
\]
and integrating both sides gives
\[
\frac{1}{8\Phi(t,x)^8} =t + C.
\]
Applying the initial condition $\Phi(0,x)=x$ and simplifying yields the explicit solution
\ba
\label{e:Phi9}
\Phi(t,x):=\frac{x}{\sqrt[8]{8 t x^8+1}},
\ea
see Example \ref{ex:steep}.

Hence, we approximate the dynamics of $X$ in two steps using the following \emph{direct splitting} scheme:
\ba
\label{e:Xn-exa}
X_0^h&:=x,\\
Y^h_{(k+1)h}&:=X^h_{kh} + Z_{(k+1)h} - Z_{kh},\\
X^h_{(k+1)h}&:=\Phi(h, Y^h_{(k+1)h})=\frac{Y^h_{(k+1)h}}{\sqrt[8]{8 h (Y^h_{(k+1)h})^8+1}}
,\quad k\in\bN_0.
\ea
The order of these operations is essential. Since the function $x\mapsto \Phi(h,x)$ is
bounded by $K_h:=1/\sqrt[8]{8h}$ for $h\in(0,1]$, the scheme $\{X^h_{kh}\}$ is bounded with probability one
and possesses finite moments of all orders.

\begin{figure*}
\centerline{\includegraphics{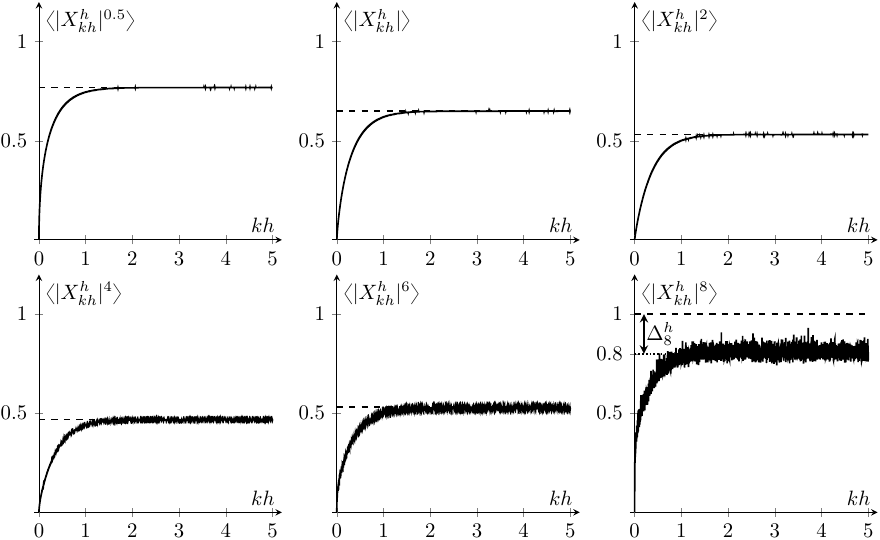}}
\caption{Empirical moments of the direct splitting scheme $X^h$ defined in \eqref{e:Xn-exa} for $h=10^{-5}$, $N=10^6$, $x=0$.
The dashed lines represent the corresponding moments of the limit distribution $X_\infty$.
\label{f:SP}}
\end{figure*}

Moreover, all moments of order $q\in (0,9)$ of the scheme converge to the corresponding moments
of the process $X$. More precisely, there is a convergence rate $\gamma\in (0,\infty)$ such that for any $T\in[0,\infty)$
\[
\sup_{kh\in[0, T]}\E |X^h_{kh}-X_{kh}|^q=\mathcal{O}(h^\gamma),\quad h\to 0,
\]
see statements 2, 3, and 4 in Appendix \ref{a:general} for a precise formulation. This result is illustrated
in Fig.~\ref{f:SP}, where we observe exponentially fast convergence
of empirical moments $\langle |X^h_{kh}|^q\rangle$
to the corresponding moments of the limit distribution $X_\infty$,
with the variance of the approximations increasing as the order
$q$ approaches the critical value $9$.

Whereas the convergence of moments up to $q=6$ is very accurate, the 8th absolute moment appears
to be underestimated. This phenomenon can be explained as follows.

Recall that the scheme $X^h$ is bounded, $|X^h_{kh}|\leq K_h$, for all $k\in\bN$, so the probability distribution of each random
variable
$X^h_{kh}$ is supported on the interval $[-K_h,K_h]$.
In other words,
the scheme $X^h$ \emph{practically} never attains values that exceed the threshold $K_h$, and thus effectively approximates
the \emph{truncated} moments of the true solution. Therefore, for large $kh$ we have
\begin{equation*}
\begin{aligned}
\langle |X_{kh}^h|^q\rangle
&\approx  \E[|X_{kh}|^q\bI(|X_{kh}|\leq K_h)]\\
&\approx  \E[|X_\infty|^q\bI(|X_\infty|\leq K_h)]=\int_{-K_h}^{K_h} |x|^q m_\infty(x)\,\di x.
\end{aligned}
\end{equation*}
Hence, the discrepancy between the true and the practically observed values of the 8th moment,
as seen in Fig.~\ref{f:SP}, is due to the difference
\[
\Delta^h_q= \E |X_\infty|^q - \E[|X_\infty|^q\bI(|X_\infty|\leq K_h)]= 2\int_{K_h}^\infty x^q m_\infty(x)\,\di x.
\]
For $h=10^{-5}$, the error $\Delta^h_q$ has the following values:
\begin{center}
\begin{tabular}{|l|c|c|c|c|c|c|c|}
\hline
$q$          & 0.5 & 1 & 2 & 4 & 6 & 7 & 8\\
\hline
$\Delta^h_q$ & $3.4\cdot 10^{-6}$ & $6.6\cdot 10^{-6}$ & $2.4\cdot 10^{-5}$  & $0.0004$  & $0.006$
&$0.03$ & $0.2$\\
\hline
\end{tabular}
\end{center}
In Fig.~\ref{f:SP}, we observe that the empirical 8th moment of $X^h$ indeed converges to $\E|X_\infty|^q-\Delta^h_8\approx 0.8$.
Furthermore, it is clear that the error term
$\Delta^h_q\to 0$ as $h\to 0$; however, this convergence becomes very slow for $q$ close to the critical
value $q=9$. Therefore, the systematic error $\Delta^h_q$ must be taken into account in numerical experiments.

\begin{figure}
\centerline{\includegraphics{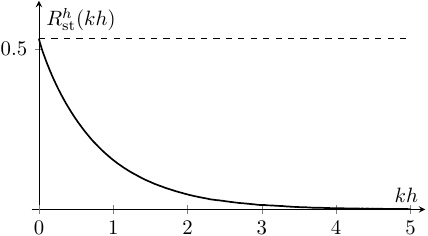}}
\caption{The estimate of the stationary autocorrelation function of the process $X$. It coincides, up to line thickness,
with the exponential function $\E_\textrm{st} X_{0}^2 \cdot\ex^{-t/\tau}= 0.532\cdot \ex^{-t/0.803}$;
$h=10^{-5}$, $N=10^6$.
\label{f:auto}}
\end{figure}

Along with individual moments, the direct splitting scheme also correctly reproduces mixed
moment characteristics of the process $X$. This can be illustrated by examining the \emph{stationary autocorrelation function}
\[
R_\text{st}(t):= \E_\textrm{st} X_{0}X_{t},
\]
where the expectation is taken under assumption that the initial value $X_0$ is distributed
according to the stationary law \eqref{e:m}, see Section 3.8 in \cite{Gardiner-04}. Simulations of the moments of
$X$ presented in Fig.\ \ref{f:SP} show that
the process $X^h$, initialized with $X_0^h=0$, reaches the stationary regime for $kh\geq 3$. Consequently, the approximation of
the stationary
autocorrelation function is obtained as
\[
R_\textrm{st}^h(kh):= \langle X^{h}_{3}X^{h}_{3+kh}\rangle ,\quad k\in\mathbb N_0,
\]
as illustrated in Fig.~\ref{f:auto}.
The \emph{correlation time} $\tau$
defined by
\[
\tau=\frac{1}{\E_\textrm{st} X_{0}^2}\int_0^\infty R_\textrm{st}(t)\,\di t
\]
can be effectively estimated as
\[
\tau^h
\approx
\frac{1}{\langle |X_3^h|^2\rangle}\sum_{k\colon kh\leq 5} R_\textrm{st}^h(kh)\cdot h= 0.803\dots
\]

\begin{figure}
\centerline{\includegraphics{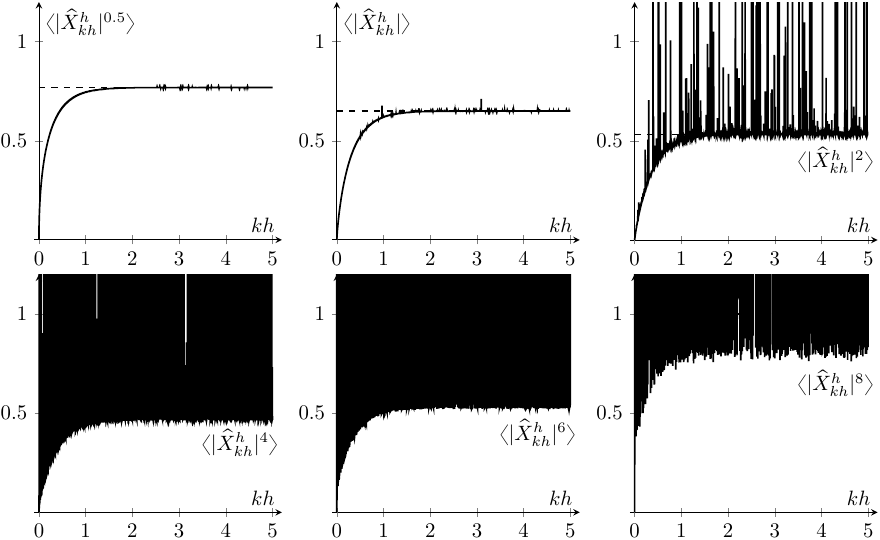}}
\caption{Empirical moments of the reverse splitting scheme $\widehat X^h$ defined
in \eqref{e:Xn-exarev} for $h=10^{-5}$, $N=10^6$, $x=0$.
The dashed lines represent the corresponding moments of the limit distribution $X_\infty$.
\label{f:RS}}
\end{figure}

We note that the order of operators $\cA$ and $\cD$ in the classical Lie--Trotter approximations is arbitrary, and we
can also consider the \emph{reverse splitting} approximations
\[
P_t \approx \widehat P^{h}_t=(\ex^{h\cD}\ex^{h\cA})^k,
\]
that correspond to the \emph{reverse splitting} scheme
\ba
\label{e:Xn-exarev}
\widehat X_0^h&:=x,\\
\widehat Y^h_{(k+1)h}&:= \Phi(h,\widehat X^h_{kh}),\\
\widehat X^h_{(k+1)h}&:= \widehat Y^h_{(k+1)h} +Z_{(k+1)h} - Z_{kh},\quad k\in\mathbb N_0.
\ea
It is clear that the scheme \eqref{e:Xn-exarev} has the same structure as the tamed Euler scheme
\eqref{e:TE-exa}, namely, it is given by a sum of a
bounded term $\widehat Y^h_{(k+1)h}= \Phi(h,X^h_{kh})$ and an increment $Z_{(k+1)h} - Z_{kh}$
which is integrable only up to order $q=1$. Therefore, the scheme $\widehat X^h$ cannot capture the moments
of the solution $X$ for orders $q\in [1,9)$.
The empirical absolute moments computed using the reverse splitting scheme $\widehat X^h$
are presented in Fig.~\ref{f:RS}, where they attain very large values for $q\in[1,9)$.

We note that a variant of the reverse splitting scheme \eqref{e:Xn-exarev} has been used in \cite{dybiec2007stationary},
where the authors also separated the noisy and deterministic dynamics and
cleverly combined the direct integration of the ODE $\dot \Phi=A(\Phi)$ for large $x$ with the
standard Euler approximation of the noisy part.

At the end of this section, we note that the explicit Euler scheme $X^{\mathrm{E},h}$ can be viewed as a first order
reduction of the splitting schemes $X^h$ and $\widehat X^h$. Indeed, expanding the function $\Phi$ given in \eqref{e:Phi9}
in a Taylor series yields
\[
\Phi(h,x)=x-x^9 h+\frac{9 x^{17}}{2}h^2 -\frac{51x^{25}}{2} h^3 +\dots.
\]
Therefore, neglecting  the higher order terms formally in the reverse splitting scheme \eqref{e:Xn-exarev} immediately recovers
the Euler scheme:
\[
\widehat X^h_{(k+1)h}\approx \widehat X^h_{kh} - (\widehat X^h_{kh})^9 h +Z_{(k+1)h} - Z_{kh},\quad k\in\mathbb N_0.
\]
For the direct splitting scheme \eqref{e:Xn-exa}, recovering the Euler scheme requires two approximations:
\begin{equation*}
\begin{aligned}
X^h_{(k+1)h}&\approx (X^h_{kh} +  Z_{(k+1)h} - Z_{kh}) - (X^h_{kh} + Z_{(k+1)h} - Z_{kh})^9 h\\
&\approx X^h_{kh} +  Z_{(k+1)h} - Z_{kh} - (X^h_{kh})^9 h.
\end{aligned}
\end{equation*}
Thus, one sees that the direct splitting scheme \eqref{e:Xn-exa} treats noise in a more sofisticated, nonlinear manner.

We also emphasize that the general theory guarantees convergence \emph{in probability} of all the schemes
\eqref{e:E-exa}, \eqref{e:TE-exa}, \eqref{e:Xn-exa}, and \eqref{e:Xn-exarev}.
However, among these, only the direct spitting scheme
\eqref{e:Xn-exa}  is free from explosion and successfully captures the higher moments of the process $X$.

\section{Error analysis of a linear system\label{s:linear}}

The second example focuses on the error analysis of a one-dimensional linear SDE
\ba
\label{e:OU}
\di X_t=-MX_t\,\di t+\di Z_t,\quad X_0=x\in\bR,
\ea
where the drift $A(x)=-Mx$ a (minus) gradient of the parabolic confining potential $U(x)=\frac{M}{2}x^2$,
with $M\in(0,\infty)$, and the driving process is
a symmetric $\alpha$-stable L\'evy process $Z$, $\alpha\in(0,2]$, with the characteristic function
$\E \ex^{\i\lambda Z_t}=\ex^{-t|\lambda|^\alpha}$. Note that when $\alpha=2$, $Z$ corresponds to a Brownian motion with variance
$\E Z_t^2= 2t$.

First we note that the solution $X$ of equation
\eqref{e:OU}
is the Ortnstein--Uhlenbeck type process, which has the following form:
\[
X_t=\ex^{-Mt}x + \int_0^t \ex^{-M(t-s)}\,\di Z_s.
\]
In particular, for each $h\in(0,1]$ and $k\in\mathbb N$ we have the following \emph{exact} representation of the solution:
\[
X_{kh}
=\ex^{-Mhk}x +  \sum_{j=1}^{k} \int_{(j-1)h}^{jh} \ex^{-Mhk+Ms}\,\di Z_s.
\]
To study the splitting schemes, we find the solution of the ODE $\dot \Phi= -M\Phi$ explicitly:
\[
\Phi(t,x)=\ex^{-Mt},\quad x\in\bR,\ t\in[0,\infty).
\]
Then the Euler, direct and reverse splitting numerical approximations take the following form for the given initial
value $X_{0}^{\mathrm{E},h}=X_{0}^h= \widehat X_0^h=x$ and $k\in\mathbb N_0$:
\begin{align}
\label{e:EUk}
X_{(k+1)h}^{\mathrm{E},h}& = X_{kh}^{\mathrm{E},h} - M X_{kh}^{\mathrm{E},h}\cdot h + Z_{(k+1)h}-  Z_{kh},\\
\label{e:SPk}
X_{(k+1)h}^h& = \ex^{-Mh}  (X_{kh}^h + Z_{(k+1)h}-  Z_{kh}),\\
\label{e:RSk}
\widehat X_{(k+1)h}^h& = \ex^{-Mh}\widehat X_{kh}^h + Z_{(k+1)h}-  Z_{kh}.
\end{align}
Since the approximations \eqref{e:EUk}, \eqref{e:SPk},  and \eqref{e:RSk}
are linear systems, we can find them explicitly:
\begin{equation*}
\begin{aligned}
X_{kh}^{\mathrm{E},h} &=(1-Mh)^k x + \sum_{j=1}^{k}(1-Mh)^{k-j} ( Z_{jh} - Z_{(j-1)h}) ,\\
X_{kh}^h &=\ex^{-Mhk}x + \sum_{j=1}^{k} \ex^{-M (k-j+1) h} ( Z_{jh} - Z_{(j-1)h}) ,\\
\widehat X_{kh}^h &=\ex^{-Mhk}x + \sum_{j=1}^{k}\ex^{-M (k-j) h} ( Z_{jh} - Z_{(j-1)h}).
\end{aligned}
\end{equation*}
These closed form formulae allow us to carry out the error analysis of the approximations. We begin with the direct
splitting scheme \eqref{e:SPk}. Define the global error of the direct splitting scheme at time $t=kh$ by
\[
e^{h}_{kh}:=|X_{kh} - X_{kh}^h|.
\]
Due to linearity, both the exact solution  $X$ and ist approximation $X^h$ can be decomposed into the sum
of a deterministic part and a random part.
Accordingly, the global (cumulative) error also splits into two components:
$e^{h,\mathrm{det}}$ and $e^{h,\mathrm{ran}}$.
We first observe that the direct splitting scheme reproduces the deterministic solution
$t\mapsto \ex^{Mt}x$ exactly on the discrete grid $\{kh\}_{k\in\mathbb N_0}$, so the
deterministic component of the global error vanishes:
\[
e^{h,\mathrm{det}}_{kh}= |\ex^{-Mkh}x -\ex^{-Mkh}x |\equiv 0.
\]
The random global error for the direct splitting method \eqref{e:SPk} is computed as follows:
\ba
\label{e:sum}
e^{h,\mathrm{ran}}_{kh}& = \Big|\int_0^{kh} \ex^{-M(kh-s)}\,\di Z_s - \sum_{j=1}^{k} \ex^{-M (k-j+1) h} ( Z_{jh} - Z_{(j-1)h})\Big|\\
&=\Big|\sum_{j=1}^{k} \int_{(j-1)h}^{jh} \Big(\ex^{-M(kh-s)} - \ex^{-M (k-j+1 )h} \Big)     \,\di Z_s\Big|\\
&=:\Big|\sum_{j=1}^{k} N_{k,j}^h\Big|.
\ea
All the summands $\{N_{k,j}^h\}$, $j=1,\dots,k$, on the right-hand side of \eqref{e:sum}
are independent $\alpha$-stable random variables with the characteristic function
\begin{equation*}
\begin{aligned}
\E \ex^{\i \lambda N_{k,j}^h}
& = \exp\Big( \i\lambda   \int_{(j-1)h}^{jh} (\ex^{-M k h + Ms}- \ex^{-M (k-j+1) h})\, \di Z_s\Big)\\
& = \exp\Big( -|\lambda|^\alpha  \int_{(j-1)h}^{jh} | \ex^{-M k h + Ms}- \ex^{-M (k-j+1) h}|^\alpha\, \di s\Big)\\
& = \exp\Big( -|\lambda|^\alpha h \ex^{-\alpha Mhk} \ex^{\alpha Mh(j-1)}\int_0^1 (\ex^{Mhu}-1)^\alpha\,\di u \Big),
\quad \lambda\in\bR.
\end{aligned}
\end{equation*}
In other words, each $N_{k,j}^h$ is $\alpha$-stable distributed with the scale parameter
\[
\sigma^h_{k,j} =  h^\frac{1}{\alpha} \ex^{-  Mhk} \ex^{  Mh(j-1)}\Big(\int_0^1 (\ex^{Mhu}-1)^\alpha\,\di u \Big)^\frac{1}{\alpha},
\]
so the random global error of the direct splitting method at time $kh$ equals the absolute value of the $\alpha$-stable random variable
with the scale parameter
\begin{equation*}
\begin{aligned}
\sigma^h_{kh} & = \Big(\sum_{j=1}^k (\sigma^h_{k,j})^\alpha\Big)^{\frac{1}{\alpha}}
=h^\frac{1}{\alpha} \ex^{-Mhk} \Big(\frac{\ex^{\alpha Mhk } -1}{\ex^{\alpha Mh} -1 }\Big)^{\frac{1}{\alpha}}
\Big(\int_0^1 (\ex^{Mhu}-1)^\alpha\,\di u \Big)^\frac{1}{\alpha},\\
e^{h,\mathrm{ran}}_{kh} &\stackrel{\di}{=}
\sigma^h_{kh} |\xi|,\quad \E\ex^{\i\lambda\xi}=\ex^{-|\lambda|^\alpha}.
\end{aligned}
\end{equation*}
For $\alpha=1$ (Cauchy noise) and $\alpha=2$ (Gaussian noise), the function
$u\mapsto (\ex^{Mhu}-1)^\alpha$ can be integrated explicitly, yielding
closed form expressions for $\sigma^h_{kh}$:
\ba
\label{e:sigma1}
\sigma^h_{kh}
&=\ex^{-Mhk} \frac{\ex^{ Mhk } -1}{\ex^{Mh} -1 } \frac{\ex^{Mh}-1-Mh}{M}
\ea
for $\alpha=1$ and
\[
\sigma^h_{kh}
=h^\frac{1}{2} \ex^{-Mhk} \Big(\frac{\ex^{2Mhk } -1}{\ex^{2Mh} -1 }\frac{\ex^{2Mh} -4\ex^{Mh} +2Mh +3}{2Mh}\Big)^{\frac{1}{2}}
\]
for $\alpha=2$.
Analogously, we obtain the global error of the reverse splitting method \eqref{e:RSk}:
\begin{equation*}
\begin{aligned}
 \widehat \sigma^h_{kh} & =h^\frac{1}{\alpha}  \ex^{-Mhk} \Big(\frac{\ex^{\alpha Mhk } -1}{\ex^{\alpha Mh} -1 }\Big)^{\frac{1}{\alpha}}
\Big(\int_0^1 (\ex^{Mh}-\ex^{Mhu})^\alpha\,\di u \Big)^\frac{1}{\alpha},\\
\widehat e^{h,\mathrm{ran}}_{kh} &\stackrel{\di}{=}
\widehat \sigma^h_{kh} |\xi|,\quad \E\ex^{\i\lambda\xi}=\ex^{-|\lambda|^\alpha},
\end{aligned}
\end{equation*}
and, in particular,
\ba
\label{e:wsigma1}
\widehat \sigma^h_{kh}
&=\ex^{-Mhk} \frac{\ex^{ Mhk } -1}{\ex^{Mh} -1 }\frac{Mh \ex^{Mh}-\ex^{Mh} +1}{M}
\ea
for $\alpha=1$ and
\[
\widehat \sigma^h_{kh}
=
h^\frac{1}{2}\ex^{-Mhk} \Big(\frac{\ex^{2Mhk } -1}{\ex^{2Mh} -1 }
\frac{\ex^{2Mh} (2Mh-3)+4 \ex^{Mh}-1}{2 Mh}\Big)^{\frac{1}{2}}
\]
for
$\alpha=2$.
The deterministic component of the global error for the Euler approximations satisfies the following relation:
\[
e^{\mathrm{E},h, \mathrm{det}}_{kh}= |\ex^{-Mhk} - (1-Mh)^k||x|.
\]
It is evident, that the Euler approximation is valid only for $Mh\in (0,1)$; in particular for $Mh\in(1,\infty)$, the deterministic
Euler approximation exhibits a sign-alternating instability.

\begin{figure}
\centerline{\includegraphics{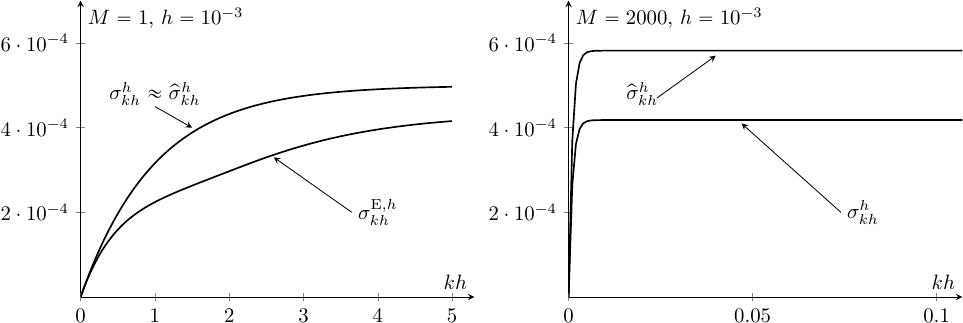}}
\caption{The scale parameters of the global random error for approximations \eqref{e:EUk}, \eqref{e:SPk},  and \eqref{e:RSk}
for $\alpha=1$ (Cauchy noise) for $M=1$ (left) and $M=2000$ (right).
\label{f:linear}}
\end{figure}

For the random component of the global error of the Euler method, we have the following equality:
\ba
\label{e:errranE}
\sigma^{\mathrm{E},h}_{kh}
&= \Big(\sum_{j=1}^{k} \int_{(j-1)h}^{jh} \Big|\ex^{-M(kh-s)} - (1-Mh)^{k-j}  \Big|^\alpha    \,\di s \Big)^\frac{1}{\alpha},
\ea
however, unfortunately, a closed-form expression for the scale parameter $\sigma^{\mathrm{E},h}_{kh}$
cannot be obtained, except in the case $\alpha=2$,
where we get
\begin{equation*}
\begin{aligned}
\sigma^{\mathrm{E},h}_{kh}
=\Big(
\frac{1-\ex^{-2 h k M}}{2 M}
-\frac{2  \ex^{-Mhk} (\ex^{Mh}-1)(\ex^{Mhk}-(1-Mh)^k)}{M (\ex^{Mh}+Mh-1)}
+\frac{1-(1-Mh)^{2 k}}{M (2-Mh)}
\Big)^{\frac12}.
\end{aligned}
\end{equation*}
The approximation errors exhibit different behaviors as functions of $kh$ depending on whether $M$ is small or large.
In Fig.\ \ref{f:linear}, we compare the scale parameters of the global errors for $\alpha=1$, with step size $h=0.001$,
under two confinement regimes: a moderate confinement $M=1$ and a strong confinement $M=2000$. In the latter case, the
original SDE \eqref{e:OU} is ill-conditioned in the computational sense, i.e., \emph{stiff}, see \cite{miranker1981numerical}
and \cite[Chapter 10.1]{kuehn2015multiple}.

For $M=1$ the scale parameters $\sigma^h$ and $\widehat\sigma^h$ are
computed using the formulae \eqref{e:sigma1} and \eqref{e:wsigma1}, respectively, while the scale parameter
$\sigma^{\mathrm{E},h}$ is obtained numerically from the formula \eqref{e:errranE}.
In all three cases, the error scale parameters exhibit a comparable order of magnitude,
$\mathcal O(h)$.

In the stiff regime,  $M=2000$, the Euler method fails to converge, whereas the global errors of the splitting schemes
stabilize exponentially fast about a level of order $\mathcal O(h)$.

In summary, the splitting
schemes should be preferred as the effective choice for simulating systems in the stiff regime.

\section{The direct splitting scheme in a general setting\label{s:general}}

Finally, we provide a formulation of the direct splitting scheme \eqref{e:Xn-exa}, \eqref{e:SPk} in a general setting.

We solve numerically a $d$-dimensional SDE with multiplicative noise:
\ba
\label{e:1}
\di X_t&=A(X_t)\,\di t+ a(X_t)\,\di t + b(X_t)\,\di B_t +  c(X_{t-})\,\di Z_t,\\
X_0&=x\in\bR^d.
\ea
The drift term in \eqref{e:1} is represented by a sum $A+a$, where
the function $A$ is $C^2$-smooth, one-sided Lipshitz continuous and
$\kappa$-confining for $\kappa\in (1,\infty)$. In the one-dimensional case,
$A$ can be interpreted as the negative gradient of a confining potential
$U$, that is,
\[
A(x)=-U'(x).
\]
The function
$a$ is assumed to be bounded and Lipschitz continuous, representing an additional regular perturbation of the confining drift,
see Examples \ref{ex:asym} and \ref{ex:rough} with $a(x)=\text{Const}$ and $a(x)=\sin x$.

The process $B=(B_t)_{t\in[0,\infty)}$ is the standard $d$-dimensional Brownian motion, and
$Z=(Z_t)_{t\in[0,\infty)}$ is an independent $d$-dimensional pure jump L\'evy process with the characteristic function
\begin{equation*}
\begin{aligned}
\E \ex^{\i \langle Z_t, \lambda\rangle}
=\exp\Big(t\int_{\bR^d}(\ex^{\i\langle \lambda,z\rangle} -1 -\i \langle \lambda, z\rangle \bI(\|z\|\leq 1)\,\nu(\di z)\Big),
\quad \lambda\in \bR^d,\quad t\in[0,\infty).
\end{aligned}
\end{equation*}
We assume that for some $p\in(0,\infty)$
the L\'evy process $Z$ has a finite $p$-th moment, $\E|Z_t|^p<\infty$, $t\in[0,\infty)$.

The  coefficients $b$ and $c$ parametrizing the multiplicative noise
are bounded and Lipschitz continuous.

Our \emph{direct splitting numerical scheme} has the following form. Let $\Phi=\Phi(t,x)$ be the solution of the ODE
\ba
\label{e:ODE}
\dot\Phi(t,x)&=A(\Phi(t,x)),\\
\Phi(0,x)&=x.
\ea
Then, we set
\ba
\label{e:Xndisc}
X_0^h&:=x,\\
Y^h_{(k+1)h}&:=X^h_{kh} + a(X^h_{kh})h + b(X^h_{kh})( B_{(k+1)h} - B_{kh})
+ c(X^h_{kh})( Z_{(k+1)h} - Z_{kh}),\\
X^h_{(k+1)h}&:= \Phi(h,  Y^h_{(k+1)h}),\quad k\in\bN_0.
\ea
In other words, the value $X^h_{(k+1)h}$ is computed in two steps.
At the intermediate step, we evaluate
the Euler--Maruyama approximation  $Y^h_{(k+1)h}$ of the random dynamics with bounded coefficients. In the second step, we
apply the deterministic confining
flow $\Phi$ to this intermediate value.

The numerical scheme \eqref{e:Xndisc} exhibits the following tail-preserving behaviour.
Under some additional technical assumptions that can be found in Appendix \ref{a:general},
for any $q\in (0,p+\kappa-1)$ the convergence
\[
\sup_{kh\in[0, T]} \E \|X_{kh}^h-X_{kh}\|^q = \mathcal {O}(h^\gamma),\quad h\to 0,
\]
holds,
with a certain positive convergence order $\gamma$.

A detailed mathematical analysis of convergence is available in \cite{aryasova2025tail}. We refer the reader to
Appendix \ref{a:general} for the complete set of mathematical assumptions and the analysis of the convergence rates.

Finally, we emphasize that the direct splitting scheme \eqref{e:Xndisc} can be equally well
applied to Gaussian SDEs (i.e., for $c(\cdot)\equiv 0$). In this case, due to the action of the confining drift
the marginal distributions of the process $X$ become \emph{sub-Gaussian}, i.e., their probability tails are lighter than Gaussian
ones, see Chapter 1 in \cite{buldygin2000metric}.
The scheme \eqref{e:Xndisc} effectively captures these very light tails, see Appendix \ref{a:general}.

\section{Ready-to-use examples\label{s:examples}}

In this section, we provide several explicit formulae for the function $\Phi$ defined in \eqref{e:ODE}.

\begin{exa}[Parabolic potential]
Let $M$ be a real $d\times d$-matrix, and define the linear drift $A(x)=-Mx$ which corresponds to the parabolic potential
$U(x)=\frac12 \langle M x,x\rangle$.
Then, $A$ is clearly Lipschitz continuous. Moreover, if all eigenvalues of $M$ have positive real parts, then $A$ is confining.
The associated linear ODE $\dot \Phi=-M\Phi$ admits the explicit solution $\Phi(t,x)=\ex^{-Mt}x$, where the matrix exponential
can be computed using standard linear algebra techniques.

This type of random dynamics describes, for example, the motion of a random charged particle in an external magnetic field;
see \cite{ChechkinGonchar:2000,ChechkinGS02}.

The advantages of the splitting schemes over the Euler method in the linear (globally Lipschitz case) have been already
discussed in Section \ref{s:linear}.
\end{exa}

\begin{exa}[Symmetric steep one-well potential, $d=1$]
\label{ex:steep}
For $c\in(0,\infty)$ and $\kappa\in (2,\infty)$, let $U(x)=\frac{c}{\kappa+1}|x|^{1+\kappa}$, so that
\[
A(x)=-U'(x)=-c|x|^{\kappa}\operatorname{sign}(x).
\]
In this case, the ODE $\dot\Phi =A(\Phi )$ has the closed form solution
\[
\Phi(t,x)=
x\Big( c(\kappa-1)t |x|^{\kappa-1} +1 \Big)^{-\frac{1}{\kappa-1}},
\]
compare with \eqref{e:Phi9} for $\kappa=9$.
The condition $\kappa\in (2,\infty)$ is required for $C^2$-smoothness of $A$.

Random dynamics in such potentials have been studied, for example, in
\cite{ChechkinGKMT-02,ChechkinGKM-04,garbaczewski2020levy,ChechkinKGMT-03,DubSpa07}.
By choosing $\kappa$ large, we can effectively approximate the so-called infinite potentials; see
\cite{DenHorHae-08,garbaczewski2020brownian,kharcheva2016spectral,pogorzelec2024role,dybiec2023escape}.
\end{exa}

\begin{exa}[Symmetric steep double-well potential, $d=1$]
For $c\in(0,\infty)$ and $\kappa\in (2,\infty)$, let $U(x)=\frac{c}{\kappa+1}|x|^{1+\kappa} - \frac{c_1}{2} x^2$, so that
\[
A(x)=-c |x|^\kappa\operatorname{sign}(x) +c_1 x.
\]
If $c_1\in (0,\infty)$, $A$ is a gradient of a symmetric double-well potential with a saddle at $x=0$ and local minima at
$x=\pm \big(\frac{c_1}{c}\big)^\frac{1}{\kappa-1}$. The mapping $\Phi$ has the explicit form
\[
\Phi(t,x)
=
c_1^{\frac{1}{\kappa-1}}
x \Big(  c|x|^{\kappa-1} (1-\ex^{-c_1 t(\kappa-1)})+  c_1\ex^{-c_1 t(\kappa-1)} \Big)^{-\frac{1}{\kappa-1}}.
\]
In particular, for $\kappa=3$,
the drift $A(x)=-c x^3+ c_1 x$, $c>0$, $c_1>0$, corresponds to the one-dimensional stochastic Ginzburg--Landau equation, see Eq.\ (4.52) in
\cite{KloPla-95}. We mention here the works \cite{Ditlevsen-99b,BrockmannS-02,ChechkinGKM-04,Pavlyukevich-06b,del2025escape}.
\end{exa}

\begin{exa}[Symmetric steep one-well potential with ``quadratic'' minumum, $d=1$]
As in the previous example, we chose
$c_1\in (-\infty,0)$, so that
\[
\Phi(t,x)
=
|c_1|^\frac{1}{\kappa-1} x
\Big( c|x|^{\kappa-1}\ex^{|c_1|(\kappa-1)t }-1) +|c_1| \ex^{|c_1|(\kappa-1)t }   \Big)^{-\frac{1}{\kappa-1}}.
\]
In this context, see \cite{ChechkinGKM-05,Lindner2008diffusion,capala2019multimodal}.
\end{exa}

\begin{exa}[Asymmetric quartic one-well potential with a cubic term, $d=1$]
\label{ex:cubic}
Let $U(x)=\frac{c}{4}x^4 -\frac{c_2}{3}x^3$ with $c\in(0,\infty)$
and $c_2\neq 0$. Then, the drift is $A(x)=-cx^3 +c_2x^2$. In this case,
$\Phi$ cannot be expressed in closed form, but it can be represented using an inverse function:
\[
\Phi(t,x)=\begin{cases}
            0,\quad x=0,\\
            c_2/c,\quad x=c_2/c,\\
           \phi^{-1}(t+\phi(x)),\quad x\notin \{0,c_2/c\},
          \end{cases}
\]
where
\[
\phi(x)=\frac{c}{c_2^2}\ln\Big|\frac{x}{x-c_2/c}\Big|-\frac{1}{c_2 x}.
\]
The function $\phi$ must be treated separately on its monotonicity intervals to properly define its inverse.
\end{exa}

\begin{exa}[Asymmetric steep potentials, $d=1$]
\label{ex:asym}
Assume that
\begin{equation*}
\begin{aligned}
U(x)&=\frac{c}{\kappa+1}|x|^{1+\kappa} - \frac{c_1}{2} x^2 - c_0 x,\\
A(x)&=-c |x|^\kappa\operatorname{sign}(x) +c_1 x+c_0,
\end{aligned}
\end{equation*}
with $c_0\in\bR$. The ODE $\dot \Phi=A(\Phi)$ cannot be solved explicitly for $c_0\neq 0$, but we can use our scheme
effectively assigning $A(x):=-c|x|^\kappa\operatorname{sign}(x) +c_1 x$ and $a(x):=c_0$.
This setting is particularly useful for analyzing bifurcations of the solution $X$ and stochastic resonance;
see \cite{ChechkinKGMT-03,DybGud09,Xu2013levy,KuhPav-17,layritz2025early}.
The same method can be applied to the drift $A(x)=-cx^3 +c_2x^2 + c_0$, see Example \ref{ex:cubic}.
\end{exa}

\begin{exa}[Nonlinear friction models]
Consider a random motion of a one-dimensional particle subject to a non-linear friction force
$A(v)$ that depends on the particle's velocity $v$, see \cite{ChechkinGKM-05,Lindner2008diffusion,EonGra-15,KP-19}.
Its dynamics is given by the kinetic Langevin equation
\begin{equation*}
\begin{aligned}
\dot x&=v,\\
\dot v&= A(v) + \mathrm{noise}.
\end{aligned}
\end{equation*}
Clearly, the drift $\boldsymbol{A}(x,y)=(v,A(v))$ of the two-dimensional system
satisfies all the assumptions that hold for $A$.
The solution of the ODE
\[
\dot{\boldsymbol{\Phi}}(t,x,v)=\boldsymbol{A}( \boldsymbol{\Phi}(t,x,v))
\]
is given by
\[
\boldsymbol{\Phi}(t,x,v)=\begin{pmatrix}
                    \displaystyle    x+\int_0^t \Phi(s,v)\,\di s\\
                        \Phi(t,v)
                       \end{pmatrix}.
\]
For the power drift $A(v)=-c|v|^{\kappa}\operatorname{sign}(v)$ from Example 2 we find that
\begin{equation*}
\begin{aligned}
 x+\int_0^t \Phi(s,v)\,\di s
= x + \frac{|v|^{2-\kappa}}{c (\kappa -2)}
\Big((c t (\kappa -1) |v|^{\kappa -1}+1)^{\frac{\kappa -2}{\kappa -1}}-1\Big)\operatorname{sign}(v).
\end{aligned}
\end{equation*}
\end{exa}

\begin{exa}[Random motion in a rough potential, $d=1$]
\label{ex:rough}
Here it is assumed that the drift $A+a$ is a gradient of a potential $U$
\[
U(x)=U_0(x)+ U_1(x),
\]
where $U_0$ is a quadratic or steep potential of the type considered in the previous examples, and $U_1$ is an
oscillatory perturbation composed of terms like $\cos(\omega x)$, $\sin(\omega x)$, with various frequences $\omega$,
see
\cite{zwanzig1988diffusion,li2016levy,dybiec2023escape}.
In this case we set $A(x)=-U'_0(x)$ and $a(x)=-U'_1(x)$, and then apply the direct splitting scheme accordingly.
\end{exa}

\begin{exa}[Rotationally invariant $d$-dimensional potential]
Let $U$ be a steep one-dimensional potential, so
that $\boldsymbol{U}(x)=U(\|x\|)$ is a steep rotationally invariant potential, $x\in\bR^d$.
Let
\[
\boldsymbol{A}(x)= -\nabla U(\|x\|)= -(U'(\|x\|)x^1,\ldots,U'(\|x\|)x^d).
\]
Let $\Phi(t,r)$ be the solution of the one-dimensional ODE $\dot\Phi=-U'(\Phi)$, $\Phi(0,r)=r\in [0,\infty)$.
Then
\[
\boldsymbol{\Phi}(t,x)=\Phi(t,\|x\|)\frac{x}{\|x\|}
\]
is the deterministic flow generated by the drift $\boldsymbol{A}$.
\end{exa}

\begin{exa}[Diagonal $d$-dimensional potential]
Assume that
$\boldsymbol{A}(x)=(A_1(x_1),\dots, A_d(x_d))$.
Then
\[
\boldsymbol{\Phi}(t,x)=(\Phi_1(t,x^1),\dots, \Phi_d(t,x^d)).
\]
\end{exa}

\begin{exa}[Singling out linear drift]
In some applications, the drift can be expressed as a sum of three terms:
$A(x) + M x +a(x)$, where $A$ is a superlinearly growing confining function, $a$ is bounded, and $M$ is a constant matrix.
Clearly, the linear term $Mx$ can be absorbed into $A$, so our theory still applies. However, especially in multidimensional
systems it is not always possible to find a solution to the
ODE $\dot \phi=A(\phi) + M\phi$, while the solution of the simpler ODE $\dot \Phi=A(\Phi)$ can often be computed explicitly.
In this case, we propose applying an additional splitting to the deterministic flow.
Let $\Phi$ be the solution of the ODE $\dot \Phi=A(\Phi)$ and $\Psi(t,x)=\ex^{Mt}x$ be the solution of the linear ODE
$\dot\Psi=M\Psi$.
Then we modify our scheme to obtain:
\ba
\label{e:Xspsp}
X_0^h&:=x,\\
Y^h_{(k+1)h}&:=X^h_{kh} + a(X^h_{kh})h + b(X^h_{kh})( B_{(k+1)h} - B_{kh})
+ c(X^h_{kh})( Z_{(k+1)h} - Z_{kh}),\\
X^h_{(k+1)h}&:= \ex^{Mh}\Phi(h,  Y^h_{(k+1)h}),\quad k\in \bN_0,
\ea
or
\ba
\label{e:Xspsp1}
X^h_{(k+1)h}&:= \Phi(h,  \ex^{Mh}Y^h_{(k+1)h}),\quad k\in \bN_0.
\ea
In both cases, it is essential that the random intermediate step $Y^h_{(k+1)h}$ is evaluated before applying the flow $\Phi$.
\end{exa}

\begin{exa}[Dissipative synchronization]
The splitting schemes \eqref{e:Xspsp} and \eqref{e:Xspsp1}
from Example 11 can be used to simulate a two-dimensional system exhibiting dissipative synchronization,
see
\cite{caraballo2008synchronization,liu2010synchronizationLevy}, given by the equation
\begin{equation*}
\begin{aligned}
\dot x = A_1(x) +\mu (y-x) + \mathrm{noise},\\
\dot y = A_2(y) +\mu (x-y) + \mathrm{noise}.
\end{aligned}
\end{equation*}
The functions $A_1$ and $A_2$ here are (minus) gradients of steep potentials as in Examples 2.\ or 4.,
and the linear term is given by the matrix
\[
M=\begin{pmatrix}
  -\mu &\mu\\
  \mu&-\mu
  \end{pmatrix},\quad \mu\in (0,\infty).
\]
The matrix exponential can be computed explicitly as
\[
\ex^{Mt}=\frac12
\begin{pmatrix}
1+ \ex^{-2 \mu t} & 1-\ex^{-2 \mu t} \\
1-\ex^{-2 \mu t} &1+\ex^{-2 \mu t}\\
  \end{pmatrix}.
\]
\end{exa}

\begin{exa}[van der Pol oscillator in Li\'enard variables]
The splitting scheme described in Example 11 also applies to the perturbed van der Pol oscillator written in Li\'enard variables
(see, e.g., Chapter 2 in \cite{gros2008complex})
as
\begin{equation*}
\begin{aligned}
\dot x &=  \mu\Big(y+ x-\frac{x^3}{3}\Big) + \mathrm{noise},\\
\dot y &= -\frac{x}{\mu} + \mathrm{noise},\quad \mu\in (0,\infty).
\end{aligned}
\end{equation*}
In this case, it is convenient to split the drift as $A_1(x,y)=\mu x-\mu x^3/3$ and $A_2(x,y)=0$, and to set
\[
M=\begin{pmatrix}
  0 &\mu\\
  -\mu^{-1}& 0
  \end{pmatrix}.
\]
This yields the following expressions for the deterministic flow $\boldsymbol{\Phi}(t,x,y)$ and the matrix exponential $\ex^{Mt}$:
\[
\boldsymbol{\Phi}(t,x,y)=\begin{pmatrix}
\frac{x}{\sqrt{x^2 (1-\ex^{-2 \mu t})/3+  \ex^{-2\mu t}}},\\
      y + t
                     \end{pmatrix}
\]
and
\[
\ex^{Mt}= \begin{pmatrix}
 \cos t & \mu  \sin t \\
 -\mu^{-1} \sin t & \cos t \\
  \end{pmatrix}.
\]
\end{exa}

\begin{exa}[FitzHugh--Nagumo system]
Analogously we can treat the FitzHugh--Nagumo system (see \cite[Chapter 1.4]{kuehn2015multiple}, \cite{buckwar2022splitting}:
\begin{equation*}
\begin{aligned}
\dot x &=  \frac{1}{\e}\Big(y+ x-\frac{x^3}{3} + \alpha \Big) + \mathrm{noise},\\
\dot y &= \gamma x - y +\beta + \mathrm{noise},\quad \mu\in (0,\infty), \ \alpha,\beta,\gamma \in\bR.
\end{aligned}
\end{equation*}
We split the drift as $A_1(x,y)=\mu x-\mu x^3/3$ and $A_2(x,y)=-y$, and set
\[
M=\begin{pmatrix}
  0 &\e^{-1}\\
  \gamma & 0
  \end{pmatrix},
\]
and $a_1(x,y)=\mu\alpha$, $a_2(x,y)=\beta$. Then the deterministic flow is
\[
\boldsymbol{\Phi}(t,x,y)=\begin{pmatrix}
\frac{x}{\sqrt{x^2 (1-\ex^{-2t/\e})/3+  \ex^{-2t\e}}},\\
      y\ex^{-t}
                     \end{pmatrix}.
\]
The matrix exponential $\ex^{Mt}$ can be expressed as
\[
\ex^{Mt}= \begin{pmatrix}
 \cosh \sqrt{\gamma \e^{-1} t} & \sqrt{\gamma^{-1}\e^{-1}} \sinh \sqrt{\gamma \e^{-1} t}  \\
\sqrt{\gamma\e}  \sinh \sqrt{\gamma \e^{-1}t } &  \cosh \sqrt{\gamma \e^{-1}t}  \\
  \end{pmatrix}
\]
for $\gamma\in(0,\infty)$ and
\[
\ex^{Mt}= \begin{pmatrix}
 \cos \sqrt{|\gamma| \e^{-1} t} & \sqrt{|\gamma|^{-1}\e^{-1}} \sin \sqrt{|\gamma| \e^{-1} t}  \\
-\sqrt{|\gamma|\e}  \sin \sqrt{|\gamma| \e^{-1}t } &  \cos \sqrt{|\gamma| \e^{-1}t}  \\
  \end{pmatrix}
\]
for $\gamma\in (-\infty,0)$.
\end{exa}

\begin{exa}[Limitations]
Many well-known dynamical systems do not satisfy the assumptions on the
coefficients $A$, $a$, $b$ and $c$ formulated in the Appendix \ref{a:general}.
In particular, the one-sided Lipschitz condition \eqref{e:Lip+}, which is crucial for proving convergence
of the splitting method, is often violated in multidimensional systems.
Examples include the van der Pol oscillator (in conventional variables),
the Duffing--van der Pol oscillator, the Lorenz system, the Brusselator, and the SIR model;
see \cite[Chapter 4]{hutzenthaler2015numerical} for further details.
\end{exa}

\section{Conclusion}

In this article, we have shown that popular simulation schemes, such as the Euler and tamed Euler methods,
fail to accurately reproduce the stochastic dynamics of heavy-tailed L\'evy flights in steep confining potentials.
In particular, the Euler scheme can exhibit blow-ups, while the tamed Euler scheme fails to capture
higher-order moments of the process, such as the variance or autocorrelation function.

To address these issues, we proposed a direct explicit splitting method that is stable, non-explosive,
and correctly captures all existing moments of the stochastic process. We demonstrated the efficiency of
this approach through a series of benchmark simulations.

We also performed an error analysis for a L\'evy-driven linear system and showed that even in the linear case,
our method outperforms existing schemes, particularly for stiff systems.

Finally, we provided a range of explicit formulas enabling the application
of our method to simulate various one- and multidimensional stochastic dynamical systems frequently encountered in practice.

%
%

%
%
%
%
%

\section*{Acknowledgements}
O.A.\ acknowledges funding from the DFG project AR 1717/2-1 (548113512).
A.C.\ acknowledges funding from the BMBF project 01DK24006 PLASMA-SPIN-ENERGY.

\section*{Use of Generative-AI Tools Declaration}
The authors declare that the text of this paper was partially reviewed for spelling
and grammar using ChatGPT.

\appendix

\section{Mathematical assumptions and convergence results \label{a:general}}

In this Appendix we list the set of rigorous mathematical assumptions
on the coefficients of the SDE \eqref{e:1} under which convergence has been established in
\cite{aryasova2025tail}.

The following notation is used.
For $x,y\in \bR^d$,
$\langle x,y\rangle$ is the usual scalar product in $\mathbb R^d$ with the associated Euclidean norm
$\|x\|=\sqrt{\langle x,x\rangle}$.
By $\bR^{d\times d}$ we denote the space of square $d$-dimensional real valued matrices.
For a matrix $M\in \bR^{d\times d}$, $M^T$ denotes the
transposed matrix, and $\|M\|$ is the matrix norm defined as the square root of the sum of the squares of the elements of $M$.
For the drift term $A\colon \bR^d\to\bR^d$, $A_x$ denotes its Jacobi matrix,
and $A^k_{xx}$ is the Hesse matrix of the component $A^k$, $k=1,\dots,d$.

For real valued functions $f$, $g$, we write $f(x)\leq_C g(x)$ if there exists a constant $C\in (0,\infty)$ such that
$f(x)\leq C g(x)$ for all $x\in \bR^d$,
and we do not need this constant for a further reference. In case that functions $f$, $g$
depend on additional parameters $t,k,h$, etc, the above inequality holds uniformly over these parameters.

\medskip

\noindent
$\mathbf{H}^\text{diss}_A$: The drift term $A$ is confining/dissipative and increases super-linearly, i.e.,
$A\colon \bR^d\to\bR^d$ is locally Lipschitz continuous
and there are $\kappa\in(1,\infty)$,  $C\in(0,\infty)$ and $C_\text{diss}\in(0,\infty)$ such that
\[
\langle A(x),x\rangle \leq -C_\text{diss}\|x\|^{1+\varkappa}+C,\quad x\in\bR^d.
\]

\smallskip

\noindent
$\mathbf{H}_{a,b,c}^{\mathrm{Lip}_b}$: The functions $a\colon \bR^d\to\bR^d$,
$b\colon \bR^d\to\bR^{d\times d}$ and $c\colon \bR^d\to\bR^{d\times d}$
are bounded and globally Lipschitz continuous.

\noindent
$\mathbf{H}_{\nu,p}$:
There is $p\in(0,\infty)$ such that the L\'evy (jump) measure $\nu$ of $Z$ satisfies
\[
\int_{\|z\|>1}\|z\|^p\,\nu(\di z)<\infty.
\]

\noindent
$\mathbf{H}_A^{\text{Lip}_+}$ The function $A\colon \bR^d\to\bR^d$ is continuous
and satisfies the one-sided Lipschitz condition, i.e.,
there is $L\in (0,\infty)$ such that
\ba
\label{e:Lip+}
\langle A(x)-A(y),x-y\rangle \leq L \|x-y\|^2,\quad x,y\in\bR^d.
\ea
\begin{rem}
It is well known that in dimension $d=1$ condition \eqref{e:Lip+} is equivalent to the boundedness of the derivative of $A$ from above,
i.e., to the condition
\[
A'(x)\leq L,\quad x\in\bR.
\]
In higher dimensions, $d\geq 2$, \eqref{e:Lip+} is equivalent to the boundedness of the symmetrized gradient matrix of $A$ from above
in the following sense:
\[
\Big\langle \frac{A_x(x)+A_x(x)^T}{2}\phi,\phi \Big\rangle \leq L\|\phi\|^2,\quad x,\phi\in\bR^d.
\]
\end{rem}

\noindent
$\mathbf{H}_{A_x,A_{xx}}$: $A\in C^2(\bR^d,\bR^d)$.
There exists $\chi\in[0,\infty)$ such that
\[
\|A_x(x)\| \leq_C 1+\|x\|^\chi,
\]
and  there exists $\e\in(0,\infty)$ such that for all $x\in \bR^d$ and $\phi\in \bR^d$
\ba
\label{e:Axx}
\e\|\phi\| \sum_{k=1}^d \|A^k_{xx}(x)\| |\phi^k|   + \langle A_x(x) \phi,\phi\rangle
\leq_C \|\phi\|^2.
\ea
\begin{rem}
In dimension $d=1$, condition \eqref{e:Axx} is equivalent to
\[
|A''(x)|\leq_C 1+\max\{-A'(x),0 \},\quad x\in\bR.
\]
Clearly, it is satisfied for any polynomial function $A$.
\end{rem}
The following results have been proven in \cite{aryasova2025tail}.

\noindent
1. Under Assumptions $\mathbf{H}^\text{diss}_A$, $\mathbf{H}_{a,b,c}^{\mathrm{Lip}_b}$, $\mathbf{H}_{\nu,p}$,
for any $q\in (0,p+\kappa-1)$
the solution $X$ of \eqref{e:1}
satisfies
\[
\sup_{t\in[0,\infty)}\E\|X_t\|^q\leq_C 1+\|x\|^q.
\]

\noindent
2. Under Assumptions
$\mathbf{H}^\mathrm{diss}_A$,
$\mathbf{H}_{a,b,c}^{\mathrm{Lip}_b}$,
$\mathbf{H}_{\nu,p},$
$\mathbf{H}_A^{\mathrm{Lip}_+}$, and
$\mathbf{H}_{A_x,A_{xx}}$, for any $q\in (0,p+\kappa-1)$ the numerical scheme $X^h$ defined in \eqref{e:Xndisc} satisfies
\[
\sup_{n\in\bN}\sup_{kh\in[0,\infty)} \E \|X^h_{kh} \|^q\leq_C 1+\|x\|^q,\quad x\in\bR^d.
\]

\noindent
3.
Under Assumptions
$\mathbf{H}^{\mathrm{Lip}_+}_A$,
$\mathbf{H}_{A_x,A_{xx}}$,
$\mathbf{H}_{a,b,c}^{\mathrm{Lip}_b}$, and
$\mathbf{H}_{\nu,p}$,
for any $T\in[0,\infty)$ and $q\in (0,p)$ there is convergence
\[
\sup_{kh\in[0, T]} \E \|X_{kh}^h-X_{kh}\|^q \leq_C h^{\frac{p-q}{\chi}\wedge\frac{q}{2}\wedge 1 }(1+\|x\|^p),
\]
and uniform convergence
\[
\E \sup_{kh\in[0, T]} \|X_{kh}^h-X_{kh}\|^q \leq_C h^{\frac{p-q}{\chi}\wedge\frac{q}{4}\wedge \frac12}(1+\|x\|^p).
\]

\noindent
4.
Under Assumptions $\mathbf{H}^\mathrm{diss}_A$, $\mathbf{H}_{a,b,c}^{\mathrm{Lip}_b}$,
$\mathbf{H}_{\nu,p}$, $\mathbf{H}_A^{\mathrm{Lip}_+}$,
and
$\mathbf{H}_{A_x,A_{xx}}$, for any $T\in[0,\infty)$, and any
$q\in [p, p+\kappa-1)$
and any
\[
\gamma
<\begin{cases}
\frac{p(p+\kappa-1 - q)}{(\chi+2)(\kappa-1)+\chi p},&\quad p\leq \chi+2, \\
\frac{p+\kappa-1 - q}{\kappa+\chi-1},  &\quad  p> \chi+2,
\end{cases}
\]
we have convergence
\[
\sup_{kh\in[0,T]}
 \E\|X^h_{kh}-X_{kh}\|^{q} \leq_C h^{\gamma }(1+\|x\|^{p+\kappa-1}).
\]

We continue with case of the Gaussian noise and assume in 5., 6.\ and 7.\ below, that $c(\cdot)\equiv 0$.
Let
\[
\Lambda=\frac{2C_\mathrm{diss}}{\sup_x\|b(x)\|^2}\in(0,\infty).
\]
Then we have the following results.

\noindent
5.
Under Assumptions
$\mathbf{H}^\mathrm{diss}_A$ and
$\mathbf{H}_{a,b,c}^{\mathrm{Lip}_b}$, for any $T\in[0,\infty)$ and $\lambda\in (0,\Lambda)$
\[
\sup_{t\in[0,T]}\E\Big[\ex^{\frac{\lambda}{1+\kappa}\|X_t\|^{1+\kappa}} \Big]
\leq_C \ex^{\frac{\lambda}{1+\kappa}\|x\|^{1+\kappa}}
\]
and for any $T\in[0,\infty)$ and $\lambda\in (0,\frac{\Lambda}{2})$
\[
\E\Big[\sup_{t\in[0,T]}\ex^{\frac{\lambda}{1+\kappa}\|X_t\|^{1+\kappa}} \Big]
 \leq_C  \ex^{\frac{\lambda}{1+\kappa}\|x\|^{1+\kappa}}.
\]

\noindent
6.
Under Assumptions
$\mathbf{H}^\mathrm{diss}_A$,
$\mathbf{H}_{a,b,c}^{\mathrm{Lip}_b}$,
$\mathbf{H}_A^{\mathrm{Lip}_+}$,
and
$\mathbf{H}_{A_x,A_{xx}}$ for any $T\in[0,\infty)$ and $\lambda\in (0,\Lambda)$
\[
\limsup_{n\to\infty}\sup_{t\in[0,T]}\E \Big[\ex^{\frac{\lambda}{1+\kappa}\|X^n_t\|^{\kappa+1}}  \Big]
\leq_C  \ex^{\frac{\lambda}{1+\kappa}\|x\|^{1+\kappa}}
\]
and for any $T\in[0,\infty)$ and $\lambda\in (0,\frac{\Lambda}{2})$
\[
\limsup_{n\to\infty}\E \Big[\sup_{t\in[0,T]}\ex^{\frac{\lambda}{1+\kappa}\|X^h_t\|^{\kappa+1}} \Big]
\leq_C  \ex^{\frac{\lambda}{1+\kappa}\|x\|^{1+\kappa}}.
\]

\noindent
7.
Under Assumptions
$\mathbf{H}^\mathrm{diss}_A$,
$\mathbf{H}_{a,b,c}^{\mathrm{Lip}_b}$,
$\mathbf{H}_A^{\mathrm{Lip}_+}$,
and
$\mathbf{H}_{A_x,A_{xx}}$, for any $T\in[0,\infty)$, $r\in(0,\infty)$ and $\lambda\in (0,\Lambda)$
\begin{equation*}
\begin{aligned}
\sup_{kh\in[0,T]}
\E&\Big[ \Big(\ex^{\frac{\lambda}{1+\kappa}\|X^h_{kh}\|^{1+\kappa}}
+\ex^{\frac{\lambda}{1+\kappa}\|X_{kh} \|^{1+\kappa}} \Big)\|X^h_{kh}-X_{kh}\|^r\Big]
\leq_C h^\frac{r}{2}(1+\|x\|^{r(1+\chi)})\ex^{\frac{\lambda}{1+\kappa}\|x\|^{1+\kappa}}
\end{aligned}
\end{equation*}
and for any $T\in[0,\infty)$, $r\in(0,\infty)$ and $\lambda\in (0,\frac{\Lambda}{2})$
\begin{equation*}
\begin{aligned}
\E&\Big[ \sup_{kh\in[0,T]}
 \Big(\ex^{\frac{\lambda}{1+\kappa}\|X^h_{kh}\|^{1+\kappa}}
+\ex^{\frac{\lambda}{1+\kappa}\|X_{kh} \|^{1+\kappa}} \Big)\|X^h_{kh}-X_{kh}\|^r\Big]
\leq_C h^\frac{r}{2}(1+\|x\|^{r(1+\chi)})\ex^{\frac{\lambda}{1+\kappa}\|x\|^{1+\kappa}}.
\end{aligned}
\end{equation*}

\section{Simulation of some heavy tailed L\'evy processes\label{a:simulation}}

In this Appendix we collect several methods for simulation of increments of
a heavy-tailed L\'evy process $Z$.

\begin{exa} Simulation of 1-dimensional $\alpha$-stable L\'evy processes.
One dimensional $\alpha$-stable random variables can be simulated with the help of the methods developed in
\cite{ibragimov1959unimodality,Zolotarev64,chambers1976method}, see Sections 1.7 and 3.3.3 in \cite{nolan2020univariate}.

A one-dimensional symmetric $\alpha$-stable L\'evy process $Z$ with the scale parameter
$\sigma\in (0,\infty)$ has the characteristic function
\[
\E \ex^{\i \lambda Z_t}= \ex^{-t \sigma^\alpha |\lambda|^\alpha}
\]
and can be simulated as
\[
Z_t\stackrel{\di}{=}
\sigma t^\frac{1}{\alpha}
\frac{\sin(\alpha U)}{(\cos U)^\frac{1}{\alpha}}\cdot \Big(\frac{\cos(U-\alpha U)}{E}\Big)^{\frac{1-\alpha}{\alpha}},\\
\]
and, in particular, for $\alpha=1$,
\[
Z_t\stackrel{\di}{=}
\sigma t \tan U,\\
\]
where the random variables $U$ and $E$ are independent and
\begin{equation*}
\begin{aligned}
U&\sim U \Big[-\frac{\pi}{2},\frac{\pi}{2}\Big]\quad \text{ and }\quad E\sim \mathrm{Exp}(1).
\end{aligned}
\end{equation*}
In general, the characteristic function of $Z_t$ (in the so-called 1-parametrization) has the form
\[
\E \ex^{\i \lambda Z_t}
=\begin{cases}
\displaystyle
\exp\big(-t \sigma^\alpha |\lambda |^\alpha\big[1- \i\beta\tan(\frac{\pi \alpha}{2})\operatorname{sign}\lambda  \big]  \big),
\quad \alpha\in (0,1)\cup(1,2),\\
\displaystyle
\exp\big(-t \sigma |\lambda | \big[1+\i\beta\frac{2}{\pi}\ln|\lambda | \operatorname{sign}\lambda\big] \big),
\quad \alpha=1.
\end{cases}
\]
For $\alpha\neq 1$ and $\beta\in[-1,1]$ we define
\[
\theta_0:= \frac{\arctan(\beta\tan(\frac{\alpha\pi}{2}))}{\alpha}
\]
and we get
\ba
\label{e:Zasym}
Z_t\stackrel{\di}{=}
\begin{cases}
\displaystyle
\frac{\sigma t^{\frac{1}{\alpha}}\sin(\alpha(U+\theta_0))} {(\cos(\alpha\theta_0)\cdot \cos(U))^{\frac{1}{\alpha}}}
\cdot
\Big(\frac{\cos(\alpha \theta_0 + (\alpha-1)U  ))}{E} \Big)^{\frac{1-\alpha}{\alpha}},
\quad \alpha\in (0,1)\cup(1,2),\\
\displaystyle
\frac{2\sigma t }{\pi}
\Big[
\Big( \frac{\pi}2 +\beta U  \Big)\tan U -\beta \ln\Big( \frac{\frac{\pi}{2}E \cos U}{\frac{\pi}2 +\beta U} \Big)
\Big],
\quad \alpha=1.
\end{cases}
\ea
\end{exa}

\begin{exa}
Simulation of $1$-dimensional compound Poisson processes with jumps sizes distributed according to the Pareto law.
In this case, the L\'evy process $Z$ has the characteristic function
\[
\E \ex^{\i \lambda Z_t }=\exp\Big(t \int_\bR (\ex^{\i \lambda z}-1)\,\nu(\di z)\Big),\quad \lambda\in\bR,
\]
where
\[
\nu(\di z)=\Big(\frac{c_-}{|z|^{1+\alpha_-}}\bI(z\leq -\sigma_-) +\frac{c_+}{z^{1+\alpha_+}}\bI(z\geq \sigma_+)\Big)\,\di z,\\
\]
with $\sigma_\pm\in (0,\infty)$, $c_\pm\in [0,\infty)$, $c_+^2+c_-^2>0$, $\alpha_\pm\in (0,\infty)$. The process $Z$ is represented as
\[
Z_t=\sum_{k=1}^{N_t} J_k,
\]
where $N_t\sim\mathrm{Poisson}(c t)$ with the
intensity
\[
c=\int_\bR\nu(\di z)=\frac{c_-}{\alpha_-\sigma_-^{\alpha_-}} + \frac{c_+}{\alpha_+\sigma_+^{\alpha_+}}\in(0,\infty),
\]
and independent jumps $\{J_k\}_{k\in\bN}$. The jumps $J_k\sim J$ are simulated as
\[
J=\begin{cases}
   -J_-,\quad \text{ with probability } p_-=\frac{c_-}{\lambda \alpha_- \sigma_-^{\alpha_-}},\\
   J_+,\quad \text{ with probability } p_+\frac{c_+}{\lambda \alpha_+ \sigma_+^{\alpha_+}},\\
  \end{cases}
\]
with $J_\pm$ being Pareto distributed with parameters $(\sigma_\pm,\alpha_\pm)$, i.e.,
\[
\P(J_\pm> u)=\Big(\frac{\sigma_\pm}{u}\Big)^{\alpha_\pm},\quad u\geq \sigma_\pm.
\]
They can be obtained with the help of a uniformly distributed random variable $U$ as
\[
J_\pm=\frac{\sigma_\pm}{U^{1/\alpha_\pm}},\quad U\sim U[0,1].
\]
\end{exa}

\begin{exa}
Simulation of $d$-dimensional isotropic or elliptically contoured $\alpha$-stable L\'evy processes.
An isotropic $d$-dimensional $\alpha$-stable random L\'evy process $Z$ has the characteristic function
\[
\E \ex^{\i \langle\lambda, Z_t\rangle}=\ex^{-t \sigma^\alpha \|\lambda\|^\alpha},
\quad \lambda\in\bR^d,\ \sigma\in (0,\infty),\ \alpha\in(0,2).
\]
It can be simulated as a time changed (subordinated) standard $d$-dimensional Brownian motion. The following explicit
formulae hold, see \cite{Pavlyukevich-07,nolan2013multivariate}. Let $W\sim\cN(0,\mathrm{Id})$ be a standard $d$-dimensional
Gaussian vector, and $S^{(\alpha/2)}$ be a standard $\frac{\alpha}2$-stable positive random variable with the characteristic function
\[
\E \ex^{\i \mu S^{(\alpha/2)}}=
\exp\big(-|\mu|^{\alpha/2}\big[1- \i\tan(\frac{\pi\alpha}{4})\operatorname{sign}\mu \big]  \big),\quad \mu\in\bR,
\]
and the Laplace transform
\[
\E \ex^{-u S^{(\alpha/2)}}=
\exp\big(-u^{\alpha/2}/\cos(\frac{\alpha\pi}{4}) \big),\quad u\in[0,\infty).
\]
According to \eqref{e:Zasym}, $S^{(\alpha/2)}$ is simulated as
\[
S^{(\alpha/2)}\stackrel{\di}{=}
\frac{\sin(\frac{\alpha}{2}V+\frac{\alpha\pi}{4})}
{(\cos(\frac{\alpha\pi}{4})\cdot \cos(V))^{\frac{2}{\alpha}}}
\cdot
\Big(\frac{\cos( (1-\frac{\alpha}{2})V -\frac{\alpha \pi}{4} )}{W}
\Big)^{\frac{2}{\alpha}-1}.
\]
Then the random vector $Z_t$ is obtained as
\[
Z_t\stackrel{\di}{=}\sigma t^{\frac{1}{\alpha}}\cdot \sqrt{2}
\Big(\cos\big(\frac{\pi\alpha}{4}\big)\Big)^{\frac{1}{\alpha}}\sqrt{S^{(\alpha/2)}}W.
\]
Applying a linear transformation $\Sigma$ to the process $Z$, we get a so-called elliptically contoured $\alpha$-stable L\'evy process,
with the characteristic function
\[
\E \ex^{\i \langle\lambda, \Sigma Z_t\rangle}=\ex^{-t \sigma^\alpha \|\Sigma^T\lambda\|^\alpha},
\quad \lambda\in\bR^d,\ \sigma\in (0,\infty),\ \alpha\in(0,2),
\]
see
\cite{nolan2013multivariate} for more details.
\end{exa}

\begin{exa}
Simulation of $d$-dimensional multifractal L\'evy processes.
Let $Z$ be given as a sum
\[
Z_t=\sum_{i=1}^N r_i Z_t^{(i)},
\]
where
$Z^{(i)}$ are independent one-dimensional heavy tail L\'evy processes with tail indices
$\alpha_i\in (0,\infty)$, and  $r_i\in\bR^d$ are non-collinear unit vectors, $\|r_i\|=1$, $\langle r_i,r_j\rangle\neq 1$, $i\neq j$.
In this case, $\alpha=\min\{\alpha_1,\ldots,\alpha_N\}$ is the heavy tail index of the heaviest component, see \cite{ImkPavSta-10}.
\end{exa}

\newcommand{\etalchar}[1]{$^{#1}$}


\end{document}